\documentclass[11pt,fleqn]{article}
\usepackage{epstopdf} 
\usepackage{graphicx}
\usepackage[a4paper,left=2cm,right=2cm]{geometry}
\usepackage{amsmath}
\usepackage{float}
\usepackage[latin1]{inputenc}
\usepackage{amsmath}
\usepackage{amsfonts}
\usepackage{amssymb}
\usepackage{float}
\usepackage[toc,page]{appendix}
\floatstyle{boxed} 
\restylefloat{figure}

\newcommand{\be}{\begin{equation}}
\newcommand{\ee}{\end{equation}}
\newcommand{\bea}{\begin{eqnarray}}
\newcommand{\eea}{\end{eqnarray}}

\begin{document}
\title{A simple model to explain the observed muon sector anomalies and small neutrino masses.}
\author{Lobsang Dhargyal\footnote{Email : dhargyal@hri.res.in} \\\\\ Harish-Chandra Research Institute, HBNI, Chhatnag Road, Jhusi Allahabad 211 019 India.}
\date{15 Nov 2017}

\maketitle
\begin{abstract}
Since its inception, no decisive departure from the predictions of Standard Model (SM) has been reported. But recently various experiments have observed few hints of possible departure from SM predictions in lepton flavor universality observables such as $R_{K^{(*)}}$, $P_{5}^{'}$, muon (g-2), $R(D^{(*)})$ etc. Many of these observable where deviation from SM in the range of (2-4)$\sigma$ were observed are related to muon ($\mu$) lepton. So these deviations may be some hint of a possible New Physics (NP) in the muon sector. In this work we extend the SM by introducing two SM singlet heavy charged leptons ($F_{e},\ F_{\mu}$) whose left handed components are charged under a new $U(1)_{F}$ gauge symmetry, one color triplet lepto-quark ($\phi_{Q}$) doublet under $SU(2)_{L}$, one inert Higgs doublet ($\phi_{l}$), three very heavy Majorana neutrinos ($N_{iR}$), all of which are odd under a $Z_{2}$ discrete symmetry. One more scalar ($\phi$) charged only under the $U(1)_{F}$ whose VEV give masses to the $U(1)_{F}$ gauge boson as well as the heavy leptons. With these new particles, we show that the observed anomalies in the muon sector as well as small neutrino masses can be explained with taking into account all the other experimental and theoretical constrains till date.
\end{abstract}

\section{\large Introduction.}

The Standard Model (SM) of particle physics turn out to be very simple but powerful mathematical construct that has stood unscathed from many experimental probe to find its loop holes for about forty years by now. Although SM itself has been verified by many experimental probes of its predictions, the neutrinos oscillation (the simplest way to interpret it, is to assume neutrinos have tiny but non-zero masses) and the dark-energy and dark-matter (DM) are clear indication of its incompleteness. But recently some intriguing anomalies has been reported by various experiments in muon (g-2) \cite{PDGg-2}, $R_{K^{(*)}}$ \cite{LHCb-B1}\cite{LHCb-B2}\cite{LHCb-B3} and $P^{'}_{5}$ \cite{LHCb-B4}\cite{LHCb-B5}\cite{LHCb-B6} which may be indications of cracks in the SM. In general the flavor changing neutral current (FCNC) are sensitive to new-physics (NP) because SM is free of FCNC at tree level. One particular FCNC mode which has been well studied is $b \rightarrow s ll$, and it provoked tremendous interest in the particle physics community when LCHb reported anomalies in $B \rightarrow K^{*}\mu^{+}\mu^{-}$, $B_{s} \rightarrow \phi\mu^{+}\mu^{-}$ and $R_{K^{(*)}} = \frac{Br(B \rightarrow K^{(*)}\mu^{+}\mu^{-})}{Br(B \rightarrow K^{(*)}e^{+}e^{-})}$ \cite{LHCb-B3}. Although the deviations in each individual modes are in the range of $(2.2 - 2.6)\sigma$, since all these mode are in the $b \rightarrow s \mu\mu$, the combine amounts to a deviation from SM prediction at 4 $\sigma$ \cite{Parnan13}. A global fit to NP indicates that a NP contributions in Wilson coefficients $C_{9}$, $C_{9} = -C_{10}$, or $C_{9} = -C_{9}^{'}$ with preference for large negative $C^{NP}_{9}$ at the level of 4-5$\sigma$ than SM \cite{Parnan12}\cite{Parnan13}\cite{Parnan14}. In this work we propose a model where NP contribute to $b \rightarrow s \mu\mu$ via box diagram to generate a NP Wilson coefficient $C_{9}^{NP} = - C_{10}^{NP}$. The combine global fit in this case for the Wilson coefficients is given as \cite{Parnan13}
\be
-0.81 \leq C_{9}^{NP} = -C_{10}^{NP} \leq -0.51\   (at\ 1\sigma).
\label{C9-C10-bound}
\ee
The most important constrains on the Wilson coefficients in these kind of models comes from the $B^{0}\ -\ \bar{B}^{0}$ mixing, $b \rightarrow s\ \gamma$ and $Br(B_{s}^{0} \rightarrow \mu\mu)$ where for the $B^{0}\ -\ \bar{B}^{0}$ mixing we have \cite{Parnan}
\be
C_{B\bar{B}}(\mu_{H})\ \epsilon\ [-2.1,0.6]\times 10^{-5} TeV^{-2}\ (at\ 2\sigma)
\label{Cbbar}
\ee
where $\mu_{H} = 2m_{W}$ and for the $Br(B_{s}^{0} \rightarrow \mu^{+}\mu^{-})$ we have \cite{Guang-Zhi-Xu2}
\be
Br(B_{s} \rightarrow \mu^{+}\mu^{-})_{Exp.} = 2.8^{+0.7}_{-0.6}\times 10^{-9}
\label{Bs-mumu}
\ee
which is 1.2$\sigma$ below the SM expectation of $Br(B_{s} \rightarrow \mu^{+}\mu^{-})_{SM} = (3.66 \pm 0.23)\times 10^{-9}$ \cite{Guang-Zhi-Xu1}. For the $b \rightarrow s\ \gamma$, the constrain on $C_{7}^{NP}$ and $C_{8}^{NP}$ at 2$\sigma$ turn out to be \cite{Parnan}
\be
-0.098 \leq C_{7}(\mu_{H}) + 0.24C_{8}(\mu_{H}) \leq 0.07\ (at\ 2\sigma),
\label{bs-gamma}
\ee
where $\mu_{H}$ is take at $2m_{W}$. Then there is also the observed anomaly in the muon (g-2)
\be
\delta a_{\mu} = a_{\mu}^{Exp.} - a^{SM}_{\mu} = 288(63)(49)\times 10^{-11},
\ee
which is at 3.6$\sigma$ deviation from SM prediction \cite{PDGg-2}.

\section{Model details.}
\label{mod-det}

In this work we would like to propose a new-physics (NP) model which can explain the anomalies observed in $R_{K^{(*)}}$ and muon (g-2) along with the loop generated neutrino masses. In line with the arguments given in \cite{Parnan}\cite{Parnan60}, we introduce two heavy scalar $\phi_{Q}$ and $\phi_{l}$, where $\phi_{Q}$ is a lepto-quark, doublet under $SU(2)_{L}$ SM gauge group, and $\phi_{l}$ is the inert-doublet of inert two-Higgs-doublet model (IDM), both of which are odd under a discrete $Z_{2}$ transformation. Three heavy charged leptons $F_{e}$, $F_{\mu}$ and $F_{\tau}$ whose left-handed components are charged under a new $U(1)_{F}$ gauge symmetry, all are odd under the discrete $Z_{2}$ transformation where the subscripts on the heavy lepton symbols denote lepton numbers they carry. We also introduce a new scalar $\phi$ charged under the $U(1)_{F}$ which develops a non-zero VEV and gives masses to the new gauge boson $Z_{\mu}^{F}$ as well as new heavy leptons $F_{e}$, $F_{\mu}$ and $F_{\tau}$. In addition to the above new particles we also add three right-handed Majorana fermions $N_{iR}$ to generate neutrino masses at one loop level \cite{Ma}. The new particles and their charge under the various transformations are tabulated in Table-\ref{tab1}. With these new particles, as pointed out in \cite{Parnan}\cite{Parnan60}, one loop box contribution to $b \rightarrow s\mu^{+}\mu^{-}$ can be generated. Due to the fact that left-handed $F_{l}$ being charged under the $U(1)_{F}$, only the right-handed components of the new particles and the left-handed components of SM fermions (which are both not charged under $U(1)_{F}$) can interact via Yukawa terms given as
\be
\mathcal{L}_{int} = \sum_{i} (y^{Q}_{i}\bar{Q}_{i}P_{R}F_{i}\phi_{Q} + y_{i}\bar{L}_{i}P_{R}F_{i}\phi_{l}) + h.c.
\label{eqt-1}
\ee

\begin{table}[h!]
\begin{center}
\begin{tabular}[b]{|c|c|c|c|c|c|} \hline
Particles & $SU(3)_{c}$ & $SU(2)_{L}$ & $U(1)_{Y}$ & $U(1)_{F}$ & $Z_{2}$ \\
\hline\hline
$\phi_{Q}$ & 3 & 2 & 7/6 & 0 & -1 \\
\hline
$\phi_{l}$ & 1 & 2 & 1/2 & 0 & -1 \\
\hline
$F_{iL}$ & 1 & 1 & $Y_{i}$ & $n_{i}$ & -1 \\
\hline
$F_{iR}$ & 1 & 1 & $Y_{i}$ & 0 & -1 \\
\hline
$N_{jR}$ & 1 & 1 & 0 & 0 & -1 \\
\hline
$\phi$ & 1 & 1 & 0 & $n_{\phi} = n_{\mu} = -n_{e}$ & +1 \\
\hline
\end{tabular}
\end{center}
\caption{The charge assignments of new particles under the full gauge group $SU(3)_{c}\times SU(2)_{L}\times U(1)_{Y}\times U(1)_{F}$. Where $i = {e, \mu, \tau}$ and $j = {1, 2, 3}$.}
\label{tab1}
\end{table}

Now the main constrains on the values of $Y_{i}$ and $n_{i}$ comes from the anomaly free conditions which gives
\be
\begin{split}
\sum^{\tau}_{i = e}Y_{i}^{2}n_{i} = 0\\
\sum^{\tau}_{i = e}n_{i}^{2}Y_{i} = 0\\
\sum^{\tau}_{i = e}n^{3}_{i} = 0
\end{split}
\label{anomaly1}
\ee
which are the anomaly free conditions coming from $U(1)_{Y}^{2}U(1)_{F}$, $U(1)_{F}^{2}U(1)_{Y}$ and $U(1)_{F}^{3}$ respectively and one more anomaly free condition due to gravity as $Gravity^{2}U(1)_{F}$ which gives
\be
\sum^{\tau}_{i = e}n_{i} = 0.
\label{anomaly2}
\ee
The simplest nontravial solution of the above four equations is given in \cite{our-paper2} by setting $n_{\tau} = 0$, then $Y_{\mu} = -Y_{e}$ and $n_{\mu} = -n_{e}$ solves the above four equations with $Y_{\tau}$ a free parameter, in this work we do without requiring the existence of $F_{\tau}$\footnote{In general with $Y_{\tau} = 0$, the $F_{\tau}$ could also be a DM candidate, but in this work we will leave that possibility for a future pursuit and if we take $Y_{\tau} = -1$, then it can contribute to $b \rightarrow s \tau \tau$ as well as $(g-2)_{\tau}$.}. We take $Y_{2} = Y_{\mu} = Q_{F_{\mu}} = -1$ then Yukawa terms given in Eqs.(\ref{eqt-1}) for $F_{1} = F_{e}$ is not allowed due to charge conservation and so no contribution to $b \rightarrow s e^{+}e^{-}$ from NP is expected, which is in line with the experimental findings that the NP is most likely in the muon sector instead of the electron sector\cite{NP-Bmumu}. We would like to point out that in \cite{our-paper2} the $F_{iR}$ is charged under the new $U(1)_{F}$ and it has been able to explain the muon (g-2) data but not $R_{K^{(*)}}$ data, in this model we realized that if we let the $F_{iL}$ to be charged under the new $U(1)_{F}$ instead, then the model will be able to explain both muon (g-2) and $R_{K^{(*)}}$ data. And also, due to charge conservation and Lorenz invariance requirement, the $F_{e}$ will be a stable heavy charged lepton\footnote{$F_{e}$ can decay with introduction of (doubly charged under $U(1)_{Y}$ and singly charged under $U(1)_{F}$) scalars with Yukawa terms (at the order of LFV in charge leptons) such as $\bar{F}_{\mu R}\phi^{--}_{-n_{e}}F_{eL}$, with the $\phi^{--}_{-n_{e}}$ having large couplings to other exotic scalars, such that its present relic density would be neglegible.} whose mass (from the latest PDG \cite{L3} lower bound for heavy charged lepton mass) is $m_{F_{e}} \geq 102.6$ GeV but searches for long lived stable charged particles (in SUSY context) at LHC put the lower bound on heavy charged leptons as $m_{F_{e}} \geq 620$ GeV \cite{Atlas620}.
Now from the Eqs.(11) of \cite{JP-Leveille} we have for the contribution of a neutral Higgs to the $\delta a_{\mu}$ coming from the Yukawa terms of Eqs.(\ref{eqt-1}) is given as
\be
\delta a_{\mu} = \frac{m^{2}_{\mu}y^{2}_{\mu}}{2\times 16\pi^{2}}\int^{1}_{0}dx[ \frac{x^{2} - x^{3}}{m_{\mu}^{2}x^{2} + (m^{2}_{F} - m^{2}_{\mu})x + m^{2}_{H^{0}_{l}}(1 - x)} + \frac{x^{2} - x^{3}}{m_{\mu}^{2}x^{2} + (m^{2}_{F} - m^{2}_{\mu})x + m^{2}_{A_{l}}(1 - x)}]
\label{delAg-2}
\ee
and in the limit $m_{A_{l}} \approx m_{F} \approx m_{H^{0}_{l}} >> m_{\mu},\ |m_{F} - m_{H_{l}}|$, we have 
\be
\delta a_{\mu} \approx \frac{m^{2}_{\mu}}{12\times 16\pi^{2}}(\frac{y_{\mu}}{m_{F}})^{2},
\ee
where if $m_{\mu} > |m_{F} - m_{H^{0}_{l}}|$ then both $F_{\mu}$ and $H^{0}_{l}$ will be stable, otherwise only the lighter of the two will be a stable particle ($H_{l}^{0}$ in our case and hence a DM candidate). But in \cite{JK}\cite{second-paper10} it has been shown that DM relic density contribution from scalar DM with Yukawa coupling in order unity (required for $H^{0}_{l}$ to explain the muon (g-2) as shown below) is negligible.
Experimentally the observed anomaly in the muon (g-2) is given as
\be
\delta a_{\mu} = a^{Exp}_{\mu} - a^{SM}_{\mu} = 288(63)(49)\times 10^{-11}
\ee
amounting to about 3.6$\sigma$ disagreement with SM prediction \cite{PDGg-2}. As first pointed out in \cite{our-paper1}, this discrepancy can be explained by a scalar and a heavy lepton ($F_{\mu}$) propagating in the loop within 1$\sigma$ of the experimental value with $\frac{y_{\mu}}{m_{F}} \approx 0.0188$, for instance setting $y_{\mu} = 3$, $m_{F_{\mu}} = 160$ GeV, $m_{H_{l}} = 150$ GeV and $m_{A_{l}} = 300$ GeV in Eqs.(\ref{delAg-2}), we get $\delta a_{\mu} = 1.751\times 10^{-9}$, which is within the 1.4$\sigma$ of measured deviation. We will use $\frac{y_{\mu}}{m_{F}} \approx 0.0188$ as a benchmark value in the following analysis. At $Y_{\mu} = 3$, $m_{F_{\mu}} = 160$ GeV, $m_{H^{0}_{l}} = 150$ GeV and $m_{A_{l}} = 300$ GeV, adopting formula II.39 of \cite{second-paper10} and also see \cite{our-paper2}\cite{our-paper1}, we get $Br(Z \rightarrow \mu^{+}\mu^{-})_{triangle} = 4.932\times 10^{-6}$ compared to the experimental average of $Br(Z \rightarrow \mu^{+}\mu^{-})_{Exp.} = (3.366 \pm 0.007)\%$ \cite{PDGg-2}, the NP contribution via triangle loop is about an order of magnitude smaller than the errors in the most precise present experimental average from PDG. Collider signature of our model are similar to those given in \cite{our-paper2}. Since the new gauge boson mass plays no role in the explainations of the anomalies in this model, the present limit of $m_{Z^{'}} > 4.5$ TeV \cite{PDGg-2} put no more constrains here.

\subsection{Scalar Sector.}
The scalar potential can be written as
\be
\begin{split}
V = \mu_{1}^{2}|H|^{2} + \mu_{2}^{2}|\phi|^{2} + \mu_{Q}^{2}|\phi_{Q}|^{2} + \mu_{l}^{2}|\phi_{l}|^{2} + \lambda_{1}|H|^{4} + \lambda_{2}|\phi|^{4} + \lambda_{Q}|\phi_{Q}|^{4} + \lambda_{l}|\phi_{l}|^{4}\\
+ \lambda_{3}|H|^{2}|\phi|^{2} + \lambda_{3QH}|H|^{2}|\phi_{Q}|^{2} + \lambda_{3lH}|H|^{2}|\phi_{l}|^{2} + \lambda_{3\phi_{Q}}|\phi_{Q}|^{2}|\phi|^{2} + \lambda_{3\phi_{l}}|\phi_{l}|^{2}|\phi|^{2}\\
+ \lambda_{3lQ}|\phi_{l}|^{2}|\phi_{Q}|^{2} + \lambda_{4Q}|H^{\dagger}\phi_{Q}|^{2} + \lambda_{4l}|H^{\dagger}\phi_{l}|^{2}\\
+ \lambda_{4Ql}|\phi_{Q}^{\dagger}\phi_{l}|^{2} + \lambda_{5l}/2[(H^{\dagger}\phi_{l})^{2} + h.c].
\end{split}
\ee
where in the unitary gauges we have $H = (0, \frac{1}{\sqrt{2}}(v + h))^{T}$, $\phi = \frac{1}{\sqrt{2}}(v_{\phi} + h_{\phi})$, $\phi_{Q} = (H_{Q}^{+5/3}, H_{Q}^{+2/3})^{T}$ and $\phi_{l} = (H^{+}_{l}, \frac{1}{\sqrt{2}}(H^{0}_{l} + iA^{0}_{l}))^{T}$. Then the masses of the scalars are given by
\be
m^{2}_{h} = \mu_{1}^{2} + 3\lambda_{1}v^{2} + \lambda_{3}v_{\phi}^{2}/2
\ee
\be
m^{2}_{h_{\phi}} = \mu_{2}^{2} + 3\lambda_{1}v_{\phi}^{2} + \lambda_{3}v^{2}/2
\ee
\be
m^{2}_{H^{+2/3}_{Q}} = m^{2}_{H^{+5/3}_{Q}} + \lambda_{4Q}v^{2}/2 = \mu_{Q}^{2} + \lambda_{3QH}v^{2}/2 + \lambda_{3\phi_{Q}}v_{\phi}^{2}/2 + \lambda_{4Q}v^{2}/2
\ee
\be
m^{2}_{H^{\pm}_{l}} = \mu_{l}^{2} + \lambda_{3lH}v^{2}/2 + \lambda_{3\phi_{l}}v_{\phi}^{2}/2
\ee
\be
m^{2}_{H^{0}_{l}} = \mu_{l}^{2} + \lambda_{3lH}v^{2}/2 + \lambda_{3\phi_{l}}v_{\phi}^{2}/2 + \lambda_{4l}v^{2}/2 + \lambda_{5l}v^{2}/2
\ee
\be
m^{2}_{A^{0}_{l}} = \mu_{l}^{2} + \lambda_{3lH}v^{2}/2 + \lambda_{3\phi_{l}}v_{\phi}^{2}/2 + \lambda_{4l}v^{2}/2 - \lambda_{5l}v^{2}/2.
\ee
Contributions from 2HDM to Peskin-Tekuchi $\Delta T$ parameter(which is the most relevent parameter in 2HDM) is $\Delta T \approx 10^{-4}$ for $m_{H^{\pm}} \approx 161$ GeV, see \cite{HJHe} for details.

\section{Anomalies and bounds on Wilson coefficients.}

In our model, which has same gauge group representation as the A-I of \cite{Parnan}, the contribution from the NP to the observed anomalies in $b \rightarrow s \mu \mu$ observables comes from the box loop, and in \cite{Parnan} the authors have done a general analysis of such models. The contribution to the Wilson coefficients of $b \rightarrow s \mu\mu$ by NP box loop is given as \cite{Parnan}
\be
C_{9}^{NP} = -C_{10}^{NP} = N\frac{y_{b}y_{s}^{*}|y_{\mu}|^{2}}{2\times 32\pi\alpha_{EM}m^{2}_{m_{F_{\mu}}}}[F(x_{Q},x_{H^{0}_{l}}) + F(x_{Q},x_{A_{l}})],
\label{C9-C10}
\ee
where $N^{-1} = \frac{4G_{F}}{\sqrt{2}}V_{tb}V_{ts}^{*}$ and
\be
F(x,y) = \frac{1}{(1-x)(1-y)} + \frac{x^{2}\ln[x]}{(1-x)^{2}(x-y)} + \frac{y^{2}\ln[y]}{(1-y)^{2}(y-x)}
\ee
with $x_{Q} = \frac{m_{\phi_{Q}}^{2}}{m^{2}_{F_{\mu}}}$, $x_{H^{0}_{l}} = \frac{m_{H^{0}_{l}}^{2}}{m^{2}_{F_{\mu}}}$ and $x_{A_{l}} = \frac{m_{A_{l}}^{2}}{m^{2}_{F_{\mu}}}$. As mentioned in section \ref{mod-det}, to explain the muon (g-2) within 1$\sigma$ of the experimental value, we need $\frac{y_{\mu}}{m_{F_{\mu}}} \approx 0.0188$, so putting this value into Eqs.(\ref{C9-C10}) and taking the benchmark values of the masses as $m_{\phi_{Q}} = 900$ GeV (which is about the present LHC lower bound \cite{Parnan91}\cite{Parnan92}), $m_{H^{0}_{l}} = 150$ GeV, $m_{A_{l}} = 300$ GeV and $m_{F_{\mu}} = 160$ GeV and setting $C_{9}^{NP} = -C_{10}^{NP} = -0.66$, which is within the 1$\sigma$ range of the present experimental bound given in Eqs.(\ref{C9-C10-bound}), we get $y_{b}y_{s}^{*} = -0.029$. At the above parameter values, we have $(C^{\gamma}_{9})_{Penguin}^{NP} << C_{9}^{NP} = -C_{10}^{NP}$. Using the benchmark values of masses and $y_{b}y_{s}^{*} = -0.029$ in
\be
C_{B\bar{B}} = \frac{(y_{b}y^{*}_{s})^{2}}{128\pi^{2}m_{F_{\mu}}^{2}}F(x_{Q},x_{Q})
\ee
we get $C_{B\bar{B}} = 7.073\times 10^{-7}$ TeV$^{-2}$ which is about an order of magnitude smaller than 2$\sigma$ present experimental bound given in Eqs.(\ref{Cbbar}) at $\mu_{H} = 2m_{W}$. And also for only two generation couplings of the new lepton, with $y_{u} = V_{us}y_{s} + V_{ub}y_{b} \approx V_{us}y_{s}$ and $y_{c} = V_{cs}y_{s} + V_{cb}y_{b} \approx V_{cb}y_{b}$ ($V_{ub} << V_{us}$ and $V_{cs} << V_{cb}$) we get $C_{D\bar{D}}^{NP} \approx 3.436\times 10^{-8} TeV^{-2}$ which is well within 2$\sigma$ experimental bound of $|C_{D\bar{D}}^{Exp}| < 2.7\times 10^{-7} TeV^{-2}$. Similarly with benchmark masses and $Y_{b}Y_{s}^{*} = -0.029$ we get
\be
C_{7}(\mu_{H}) + 0.24C_{8}(\mu_{H}) = -2.538                                   \times 10^{-3},
\ee
which is almost two-orders of magnitude smaller than the present 2$\sigma$ experimental bound on this combination of Wilson coefficients coming from $b \rightarrow s\ \gamma$ data given in Eqs.(\ref{bs-gamma}), where
\be
C_{7} = \frac{Ny_{b}y_{s}^{*}}{2m_{F_{\mu}}^{2}}[\frac{2}{3}F_{7}(x_{Q}) + \tilde{F}_{7}(x_{Q})]
\ee
and
\be
C_{8} = \frac{Ny_{b}y_{s}^{*}}{2m_{F_{\mu}}^{2}}[F_{7}(x_{Q}]
\ee
with $F_{7}(x) = \frac{x^{3}-6x^{2}+6xlog[x]+3x+2}{12(x-1)^{4}}$ and $\tilde{F}_{7}(x) = \frac{1}{x}F_{7}(x^{-1})$. Now another key observable that put very stringent constrain comes from the measurement of $Br(B_{s} \rightarrow \mu^{+}\mu^{-})_{Exp.} = 2.8^{+0.7}_{-0.6} \times 10^{-9}$ which is about 1.2$\sigma$ below the SM prediction of $Br(B_{s} \rightarrow \mu^{+}\mu^{-})_{SM} = (3.66 \pm 0.23) \times 10^{-9}$ \cite{Guang-Zhi-Xu}. The decay $B_{s} \rightarrow \mu^{+}\mu^{-}$ can be expressed as
\be
Br(B_{s} \rightarrow \mu^{+}\mu^{-})_{eff.} = \frac{G_{F}^{2}\alpha_{EM}^{2}}{16\pi^{3}}|V_{tb}V^{*}_{ts}|^{2}|C_{10}^{eff.}|^{2}m_{\mu}^{2}m_{B_{s}}f^{2}_{B_{s}}(1 + \mathcal{O}(m_{\mu}^{2}/m_{B_{s}}^{2})),
\ee
where $C_{10}^{eff.} = C_{10}^{SM} + C_{10}^{NP}$ with $C_{9, 10}^{SM} = (4.07, -4.31)$ \cite{Parnan13} and $C_{10}^{NP} = +0.66$ as our benchmark value, we get $Br(B_{s} \rightarrow \mu^{+}\mu^{-})_{eff.} = 2.63\times 10^{-9}$ which is well within the 1$\sigma$ of the measured value. The bound coming from $B \rightarrow K^{(*)}\nu \nu$ is much weaker than the experimental bounds from $B \rightarrow K^{(*)}\mu^{+} \mu^{-}$, so we can ignore constrain from this mode \cite{Parnan}.

\section{Loop generation of neutrino masses.}

With presence of $N_{jR}$ we can have Yukawa terms such as
\be
\mathcal{L}_{Y} = \sum_{i,j = 1}^{3} h_{ij} \bar{L}_{i}i\sigma_{2}\phi_{l}N_{jR} + h.c,
\ee
which is well known to give Majorana neutrino mass term $M_{\alpha\beta}\bar{\nu}_{\alpha}^{c}\nu_{\beta} + h.c$ at one loop level via the scotogenic mechanism given as \cite{Ma}
\be
M_{\alpha\beta} = \sum_{i}\frac{h_{\alpha i}h_{\beta i}M_{i}}{16\pi^{2}}[\frac{m_{H_{0}}^{2}}{m_{H_{0}}^{2} - M^{2}_{i}}\ln\frac{m_{H_{0}}^{2}}{M^{2}_{i}} - \frac{m_{A_{0}}^{2}}{m_{A_{0}}^{2} - M^{2}_{i}}\ln\frac{m_{A_{0}}^{2}}{M^{2}_{i}}],
\label{nu-mass}
\ee
where $m_{H_{0}}$ and $m_{A_{0}}$ are masses of the $H_{0}$ and $A_{0}$ respectively and $M_{i}$'s are the very heavy Majorana masses of the $N_{iR}$ neutrinos. In our benchmark values where $m_{H_{0}} = 150$ GeV and $m_{F_{\mu}} = 160$ GeV with assuming $m_{H_{0}} < m_{A_{0}} = 300$ GeV, the $H_{0}$ will be a DM candidate but due to requirement of large Yukawa coupling between $H_{0}$, $F_{\mu}$ and $\mu$ to explain the muon (g-2) data, as pointed out in \cite{second-paper10}, for such large Yukawa couplings the contribution to the present relic density of the DM by $H_{0}$ will be negligibly small. A Yukawa couplings of order $|h_{11}|^{2} \approx 10^{-7}$ with lightest of the heavy neutrino mass about $2.6\times 10^{7}$ will be able to generate neutrino masses of order $\mathcal{O}(0.04)$ eV which is close to the latest experimental bound on largest of neutrino mass difference from nuetrino mixing measurements of $|\Delta m_{32}| \approx 0.05$ eV \cite{PDGg-2}, for more details see also \cite{Ma}. From such heavy $N_{R}$ the contributions to $\delta a_{l}$, $C_{9,10}^{NP}$ etc. are negligible.

\section{Conclusions.}

In this work we have proposed a simple model which can explain the observed muon related anomalies along with small neutrino masses. We have introduced one leptoquark ($\phi_{Q}$ which is triplet under $SU(3)_{c}$) and one inert Higgs doublet $(\phi_{l})$, both are odd under a $Z_{2}$ and doublet under $SU(2)_{L}$, at least two $SU(2)_{L}$ singlet heavy leptons $F_{e}$ and $F_{\mu}$, both odd under a $Z_{2}$ and whose left handed components are charged under a new $U(1)_{F}$ gauge symmetry. One $Z_{2}$ even scalar singlet under the SM gauge groups but charged under the new $U(1)_{F}$ gauge symmetry whose VEV gives masses to the new heavy leptons and $U(1)_{F}$ gauge boson. Also we added three very heavy right handed Majorana neutrinos odd under the $Z_{2}$ to generate neutrino masses at one loop via the scotogenic mechanism.\\
\\


{\large Acknowledgments: \large} This work is supported and funded by the Department of Atomic Energy of the Government of India and by the Government of U.P.


\begin{thebibliography}{99}

\bibitem{PDGg-2} C. Patrinani et al. (Particle Data Group), \textsl{Chin. Phys. C, 40, 100001 (2016) and 2017 update and references there in.}

\bibitem{LHCb-B1} J. Matias, F. Mescia, M. Ramon and J. Virto (LHCb), \textsl{JHEP 04, 104 (2012), 1202.4266.}

\bibitem{LHCb-B2} R. Aaij et al. (LHCb), \textsl{Phys. Rev. Lett. 111 (2013) 191801.}

\bibitem{LHCb-B3} R. Aaij et al. (LHCb), \textsl{Phys. Rev. Lett. 113 (2014) 151601.}

\bibitem{LHCb-B4} S. Descotes-Genon, T. Hurth, J. Matias and J. Virto (LHCb), \textsl{JHEP 1305, 137 (2013), 1303.5794.}

\bibitem{LHCb-B5} R. Aaij et al. (LHCb), \textsl{JHEP 02, 104 (2016).}

\bibitem{LHCb-B6} A. Abdesselam et al. (Belle), \textsl{LHCSki 2016 Abergurgl, Tyrol, Austria, April 10-15, 2016 (2016).}

\bibitem{Parnan12} W. Altmannshofer and D. M. Straub, \textsl{arXiv: 1503.06199.}

\bibitem{Parnan13} S. Descotes-Genon, L. Hofer, J. Matias and J. Virto (2015), \textsl{arXiv: 1510.04239.}

\bibitem{Parnan14} T. Hurht, F. Mahmoudi and S. Neshatpour (2016), \textsl{arXiv: 1603.00865.}

\bibitem{Parnan} P. Arnan, Lars Hofer, F. Mescia and A. Crivellin (2017), \textsl{DOI: 10.1007/JHEP04(2017)043.}

\bibitem{Guang-Zhi-Xu2} V. Khachatryan et al. (LHCb, CMS), \textsl{Nature 522 (2015) 68-72.}

\bibitem{Guang-Zhi-Xu1} C. Cobeth et al., \textsl{Phys. Rev. Lett. 112 (2014) 101801.}

\bibitem{Parnan60} B Gripaios, M. Nardecchia and S. A. Renner, \textsl{JHEP 06, 083 (2016).}

\bibitem{Ma} E. Ma, \textsl{Phys. Rev. D 73, 077301 (2006).}

\bibitem{Gambit-2017} The GAMBIT collaboration, \textsl{DOI: 10.1140/epjc/s10052-017-5113-1.}

\bibitem{our-paper2} Lobsang Dhargyal, \textsl{DOI: 10.1140/epjc/s10052-018-5641-3. arXiv:1709.04452 [hep-ph].}

\bibitem{NP-Bmumu} M. Bordone, G. Isidori and A. Pattori \textsl{DOI: 10.1140/epjc/s10052-016-4274-7.}

\bibitem{L3} L3 Collaboration, \textsl{Phys. Lett. B 517 (2001) 75-85.}

\bibitem{Atlas620} The ATLAS Collaboration, \textsl{JHEP 1501 (2015) 068.}

\bibitem{JP-Leveille} J. P. Leveille, \textsl{Nuclear Physics B137 (1978) 63-76.}

\bibitem{our-paper1} Lobsang Dhargyal, \textsl{arXiv:1705.09610 [hep-ph].}

\bibitem{second-paper10} C. W. Chaing, H. Okada and E. Senaha, \textsl{Phys.Rev. D96
(2017) no.1, 015002.}

\bibitem{Guang-Zhi-Xu} Guang-Zhi Xu, Yue Qiu, Cheng-Ping Shen and Yu-Jie Zhang \textsl{arXiv: 1601.03386v2.}

\bibitem{Parnan91} ATLAS Collaboration, \textsl{Eur. Phys. J. C 76, 547 (2016).}

\bibitem{Parnan92} CMS Collaboration, \textsl{Eur. Phys. J. C 76, 439 (2016).}

\bibitem{JK} J. Kawamura, S. Okawa and Y. Omura, \textsl{arXiv:1706.04344}

\bibitem{HJHe} H.J He, N. Polonsky and S. Su, \textsl{arXiv:hep-ph/0102144}


\end{thebibliography}
\end{document}